# Tunable metasurfaces via subwavelength phase shifters with uniform amplitude

Shane Colburn[1], Alan Zhan[2], and Arka Majumdar[1,2, *]

[1] Department of Electrical Engineering, University of Washington, Seattle.

[2] Department of Physics, University of Washington, Seattle.

* Corresponding Author: arka@uw.edu

## Abstract:

Metasurfaces with tunable spatial phase functions could benefit numerous applications. Currently, most approaches to tuning rely on mechanical stretching which cannot control phase locally, or by modulating the refractive index to exploit rapid phase changes with the drawback of also modulating amplitude. Here, we propose a method to realize phase modulation at subwavelength length scales while maintaining unity amplitude. Our device is inspired by an asymmetric Fabry-Perot resonator, with pixels comprising a scattering nanopost on top of a distributed Bragg reflector, capable of providing a nearly $2\pi$ nonlinear phase shift with less than 2% refractive index modulation. Using the designed pixels, we simulate a tunable metasurface composed of an array of moderately coupled nanopost resonators, realizing axicons, vortex beam generators, and aspherical lenses with both variable focal length and in-plane scanning capability, achieving nearly diffraction-limited performance. The experimental feasibility of the proposed method is also discussed.

# Introduction:

Spatial light modulators (SLMs) are essential for many applications, including beam steering[1], holography[2,3], and microscopy[4]. Most SLMs have large pixel sizes (~10-100 optical wavelengths), which inefficiently disperse light to higher diffraction orders, and have a low refresh rate (~100 Hz), hindering real-time modulation of dynamic wavefronts. This rate is limited by usage of liquid crystals (LCs), which have a slow response time[5]. The large spatial extent of LCs also restricts downsizing devices, which hinders usage of SLMs for applications which require ultra-compact components, as in implantable microscopy[6,7]. MEMS-based modulators[8,9] can provide higher speeds, but not only are these devices challenging to design and build[10], they are also more prone to failure due to their moving parts, and with them it is challenging to provide analog phase control. Metasurfaces, ultrathin structures composed of quasiperiodic arrays of subwavelength scatterers, or optical antennas[11,12], are a promising candidate for the realization of compact and efficient SLMs. With appropriate patterning and placement of scattering elements, metasurfaces can implement arbitrary spatial transfer functions which can modify the phase, amplitude, and polarization of incident electromagnetic waves[13-16]. Recent works demonstrating static metasurface implementations of optical elements, such as blazed gratings[17-19], lenses[20-26], vortex beam generators[27-32], holographic plates[33,34], invisibility cloaks[35,36], multi-wavelength diffractive optics[37-43], and freeform optics[44] show great promise for realizing compact optical systems. A dynamic structure composed of independently operating and individually tunable subwavelength phase elements is a prerequisite for metasurface-based spatial light modulation.

Unfortunately, most of the work regarding tunable metasurfaces thus far has relied on techniques which are either power-inefficient or incapable of tuning elements individually.

While metasurfaces transferred onto stretchable substrates[45-47] have demonstrated variable focal lengths, mechanical stretching cannot tune individual elements, limiting applicability to transfer functions with symmetries related to the stretch axis. With optically controlled phase-change material implementations[48], the modulation speed is limited to that of another LC-based SLM, while electrical control of such a device is impossible because the pixels are not electrically isolated. For approaches utilizing free carrier refraction[49], large changes in amplitude occur over the phase modulation range, and those based on conducting oxides[50] face the additional challenge of small change in phase due to a small effective volume where the refractive index changes. In general, for techniques which directly modulate the refractive index of the scattering elements, it is challenging to achieve a full $2\pi$ phase modulation range when operating in a non-resonant regime due to weak thermo-optic and electro-optic effects. By operating in a resonant regime, weak light-matter interactions can be enhanced to induce large, nonlinear phase shifts by utilizing multiple roundtrips inside the resonator. Unfortunately, large changes in phase on resonance are often accompanied by large changes in amplitude. Unlike previous approaches, by exploiting a device structure inspired by an asymmetric Fabry-Perot resonator (also known as a Gires-Tournois etalon[51]), we describe a method for achieving tunable subwavelength scattering elements with uniform amplitude and 0 to $2\pi$ phase with a small change in refractive index. We show that even when each pixel consists of a single scatterer, the effect of the optical resonator is preserved, and the effective phase change is amplified.

Asymmetric Fabry-Perot-inspired modulators and phased-arrays have been explored previously[10,52,53], but have consisted of an array of grating elements per pixel or stacked high contrast gratings[54]. Both of these approaches fail to provide subwavelength spatial resolution, which is necessary for micron-scale focal lengths and high phase curvatures. Similarly, RF-inspired optical phased-arrays[55] based on Mach-Zehnder interferometers and waveguides with polysilicon heaters have also shown simultaneous amplitude and phase tunability, but with the requirement of large pixel area. Instead, we design subwavelength pixels which consist of a single scattering nanopost atop a distributed Bragg reflector (DBR). We apply this analysis to design a compact and tunable reflective metasurface and report nearly diffraction-limited focal scanning via electromagnetic simulation.

## Results:

*Design of Scatterers:*

An asymmetric Fabry-Perot cavity is a resonator consisting of a medium bound by two mirrors with different values of reflectivity (Fig. 1a). In the case of a lossless medium between the two mirrors and with a perfectly reflecting bottom mirror, a top mirror reflectivity $r$, an input E-field amplitude $A$, and a cavity roundtrip phase accumulation of $\delta$, the complex output E-field $E_{out}$ and its phase are given by (see Supplementary Equations 1-11 for detailed derivation):

$$E_{out} = \frac{A(r+e^{i\delta})}{1+re^{i\delta}}, (1) \qquad \varphi = \tan^{-1}\left[\frac{(1-r^2)\sin\delta}{2r+(r^2+1)\cos\delta}\right], (2)$$

We can easily verify that for all possible values of $r$ and $\delta$, we have $|E_{out}| = A$, the input E-field amplitude. As $\delta$ is a function of wavelength, this relationship also holds for all input frequencies, producing a uniform output amplitude spectrum; all energy incident upon the top mirror eventually reflects off and out of the cavity. While the amplitude spectrum is flat, the phase depends strongly on both $r$ and $\delta$. For different fixed values of $r$, the phase of $E_{out}$, denoted $\varphi$, is plotted as a function of $\delta$ in Fig. 1b. The output phase $\varphi$ changes rapidly around $\delta = \pi$. As $r$ increases, the change in $\varphi$ near $\delta = \pi$ also increases, and with $r$ close to unity we find that $\varphi$ changes very abruptly by nearly $2\pi$. This region of nonlinear output phase shift arises from tuning the cavity on and off resonance. With higher values of r, the width of the resonance narrows with more roundtrips in the medium. In certain regimes, cavities with lossy media can also reduce the required change in cavity phase to achieve nearly $2\pi$ output phase shifts, but for a practical device the benefit of this narrowing would likely be offset by the reduction in amplitude efficiency from material absorption (see Supplementary Equations 16-19 and Fig. S2-S4 online). The degraded performance of such a lossy cavity can also be characterized in terms of its quality factor, which decays rapidly as the attenuation increases (see Supplementary Equations 12-15 and Fig. S1 online).

To realize a phase shifter, we exploit this regime of lossless nonlinear phase change. With a high $r$, we can choose an appropriate $L$ to put $\delta$ in this nonlinear regime, and tune the value of $n$ over a small range in order to achieve 0 to $2\pi$ phase shifts. To verify this technique, we used rigorous coupled-wave analysis[56] (RCWA) to simulate a cavity at 1550 nm consisting of a 2D grating of identical cylindrical posts patterned on a slab of silicon, on top of a distributed Bragg reflector (DBR) of 10 paired layers of silicon and silicon dioxide with high reflectivity ($R \cong 1$) (Fig. 1c). The posts were of height 324 nm, diameter 750 nm, and lattice constant 850 nm, while the silicon slab had a thickness of 180 nm. Fig. 1d shows the reflection coefficient of the cavity as a function of the silicon slab's refractive index $n$. The simulated structure provides uniform amplitude and a nonlinear phase shift, corresponding well with the expected behavior from the ideal model.

To implement arbitrary spatial phase profiles, a configuration of such cavities could be patterned across a substrate, assigning the refractive index of each cavity such that the corresponding phase shift in Fig. 1d matches the desired local phase shift. With this technique, each unit cell of the device is composed of a single asymmetric Fabry-Perot resonator. This approach has been well-explored previously, in which such structures have been patterned to form beam-steering arrays and phase-only modulation of spatial light distributions[10,52-54]. While this methodology may enable implementation of arbitrary phase profiles up to the limit of the Nyquist-Shannon sampling theorem, it would not provide subwavelength resolution due to the required lateral spatial extent of the cavity. Even though the cavity design only explicitly specifies a spatial extent in the normal direction (i.e. parallel to the propagation direction) in terms of the cavity length, the lateral extent is assumed to be infinite when simulating in RCWA, as a periodic boundary condition is applied which forms a grating of infinite extent. As the grating on top of the cavity affects the reflectivity, changes in its design will alter the characteristics of the resonator. In practice, such infinite gratings can be approximated by a finite number of elements, but the lateral extent and the number of grating elements must still be sufficiently large for the scattering behavior to be similar to that of the infinite grating. As such, the lateral extent would still greatly exceed the lattice constant and would prevent implementation of phase profiles with subwavelength pixels. While we could continue to reduce the lateral extent by decreasing the number of grating elements, the performance would deviate further from the ideal behavior; however, recent works[23,57] show that when there are weak interactions between adjacent elements and when there is a dependence on the global phase distribution, a metasurface can operate even when a grating is approximated with a single element. Motivated by this observation, we explore the characteristics of a device in the limit where there is a single grating post per pixel, shown schematically in Fig. 2a. In this limit, the slab of silicon which previously formed the cavity and extended to infinity is reduced to a circular slab of silicon with diameter equal to that of one of the posts, such that the grating post and cavity are one in the same, forming their own isolated resonator. When this pixel design consisting of a single grating post on top of a DBR is incorporated into a lattice of such pixels, the result is a grating patterned directly on top of the DBR. As these pixels are of subwavelength lateral extent, metasurfaces synthesized using these pixels would provide subwavelength resolution, unlike the structure of Fig. 1c which requires an array of many grating elements for a single pixel.

The efficacy of the reduction of an infinite grating to a single post depends on the coupling between the grating elements. For high contrast metasurfaces, previous designs found weakly coupled elements by determining parameter regimes in which the phase characteristics are invariant under changes in the lattice constant. In this regime, each scatterer could be modelled as a truncated waveguide supporting multiple low quality factor Fabry-Perot resonances[23]. With our scatterers, the high reflectivity DBR mirror enhances the light-matter interactions and increases the finesse of these resonances, making it challenging to find parameter regimes where there is negligible variation in phase under changes in the lattice constant. As such, we compromised and strove for resonators that are moderately coupled (i.e. scatterers with slightly increased resonance width which still provide an abrupt phase shift, but are sufficiently weakly coupled such that we can accurately implement phase profiles for aspherical lenses, axicons, and vortex generators).

We varied both the refractive index and lattice constant to find that depending on the geometric parameters, diverse phase characteristics are achievable. For example, in the case of Fig. 2b, with the same DBR design as before, posts of height 680 nm, diameter 550 nm, and a 1550 nm input, the reflection coefficient was calculated by RCWA as the lattice constant was swept from 675 to 975 nm while the refractive index was varied from 3.4 to 3.6. With this design, rapid phase transitions occur as the post refractive index is swept, whereas in the case of Fig. 2c with posts of height 504 nm and diameter 750 nm, a more moderate transition occurs at a lattice constant of 850 nm, indicated by the dashed white line. In both cases, the phase exhibits a strong dependence on the lattice constant; however, for the design of Fig. 2c the broad width of the resonance indicates a weaker dependence relative to the narrow and highly resonant nature of the transitions in Fig. 2b. While even broader resonances with far less dependence on the lattice constant are achievable, such regimes are of little interest as they would offer little benefit in terms of providing a small refractive index range to achieve a full $2\pi$ phase shift. This presents a tradeoff between achieving a very narrow resonance which can very rapidly achieve a full $2\pi$ shift and having weakly coupled scatterers which allow implementation of high resolution phase profiles—a highly resonant scatterer would be very sensitive to perturbations to adjacent scatterers and would prevent realization of high gradient profiles, whereas a broad resonance would provide weak coupling and high resolution at the cost of having to change the refractive index over a wider range. As such, we compromise to achieve a wider, though still small, change in refractive index with reduced spatial resolution and select the moderate transition of Fig. 2c at a period of 850 nm and show a 1D slice of the phase as a function of refractive index in Fig. 2d, with the phase corrected so that it does not wrap modulo $2\pi$. This regime provides an exploitable nonlinear phase shift of nearly $2\pi$ for a small change in refractive index from 3.476 to 3.535 (< 2% change). Furthermore, over the full modulation range we achieve unity amplitude. To show the moderate nature of the coupling between the scattering posts, we calculated the magnetic energy density for off (Fig. 2e) and on (Fig. 2f) resonance cases of a periodic array of scattering elements with the parameters used in generating Fig. 2d. The incident plane wave has a magnetic energy density of unity and we see high confinement of energy within the resonators, with smaller but nontrivial energy densities between pillars, indicating a moderate level of coupling. Our reported energy densities are on the order of a magnitude higher than those found by similar methods in the design of high contrast transmitarrays of silicon nanoposts[23], indicating greater energy confinement within the grating layer and higher finesse.

*Tunable Metasurface Simulation Results:*

With the scattering post parameters of Fig. 2d, we designed an 80 $\mu m$ × 80 $\mu m$ structure consisting of an array of posts and implemented phase profiles for aspherical lenses defined by:

$$\varphi(x,y) = \frac{2\pi}{\lambda}\left(\sqrt{(x-x_0)^2 + (y-y_0)^2 + f^2} - f\right), \quad (3)$$

where $f$ is the focal length, $x$ and $y$ are the coordinates in the plane of the metasurface, $\lambda$ is the operating wavelength, and $x_0$ and $y_0$ are the in-plane shift amounts for the position of the focal spot. The phase profiles are implemented by mapping the desired phase at each point to one of ten possible refractive indices from Fig. 2d which correspond to phase points which span 0 to $2\pi$. By modulating the refractive index of each scattering element in this way, we demonstrate a device with both adjustable focal length and in-plane scanning capability (Fig. 3a) by finite-difference time-domain (FDTD) simulation near the surface and subsequent propagation to further planes using the angular spectrum method (see Supplementary Equations 25-27). For focal length adjustment, $f$ is swept from 50 $\mu m$ to 300 $\mu m$ with everything else fixed, whereas for focal scanning $x_0$ is swept from $+30$ $\mu m$ to $-30$ $\mu m$ in the 100 $\mu m$ focal plane. Tuning with such a small focal length is not possible in existing phase modulators as the large pixel area limits the spatial resolution and curvature of the achievable phase profiles, necessitating the use of subwavelength tunable pixels.

To characterize the metasurface lenses, we found the beam spot sizes in terms of their full width at half maximum[57] (FWHM) and compared to the diffraction-limited FWHM. For this calculation, a 1-D slice of the intensity profile in the focal plane for each focal spot was fit to a Gaussian function, from which the FWHM was extracted. The beam spot sizes are plotted for the focal length sweep and scanning cases in Fig. 3b and Fig. 3c respectively, and we observe that the focal spots are close to diffraction-limited. We also characterized the metasurfaces in terms of focusing efficiency and found a trend of efficiency increasing with focal length, with up to 41% efficiency at $280\ \mu m$ focal length (see Supplementary Fig. S5 online).

With the same metasurface used for realizing the lenses of Fig. 3, we also generated approximate Bessel beams by implementing axicons of the form:

$$\varphi(x, y) = \frac{2\pi}{\lambda} \sqrt{x^2 + y^2} \sin\beta, \quad (4)$$

where $\beta$ gives the axicon angle[26]. We designed and simulated axicons with $\beta = 4°$ and $\beta = 5°$ (Fig. 4a) and find substantially reduced diffraction over a large range when we excite the structure with a $30\ \mu m$ waist radius Gaussian beam. We also implemented vortex beam generators with tunable topological charge (Fig. 4b), obeying the phase profile:

$$\varphi(x, y) = \frac{2\pi}{\lambda} \left( \sqrt{x^2 + y^2 + f^2} - f \right) + l\theta, \quad (5)$$

## Discussion:

While we report nearly diffraction-limited focal spots with the designed tunable metasurface, such performance is limited to profiles with phase gradients that can be accurately sampled by the subwavelength lattice. Assuming the Nyquist-Shannon sampling criterion is already met, accurate phase sampling requires minimal coupling between adjacent nanoposts, such that the desired local phase shifts can be imparted without distorting the surrounding wavefront. As indicated previously, our structures do have moderate coupling between scatterers, and have limitations in terms of implementing arbitrary phase profiles. For example, our heuristically determined nanopost parameters lend themselves well to implementing lens and axicon phase profiles, but for more exotic designs with high phase gradients, such as those for holograms or higher order polynomial freeform optical surfaces[58], parameter regimes with even less element coupling may yield superior results. This behavior is evident in the degraded shape of our generated vortices (Fig. 4b), indicating phase modulation error introduced by the designed scatterers. By utilizing the nanopost designs of Fig. 2d to make unit cells comprising an arrangement of multiple identical nanoposts instead of a single scatterer, a broader range of achievable phase profiles is possible, including those for generating holograms (see Supplementary Fig. S8 online). This is an indication that lower phase gradients can reduce pixel-to-pixel coupling and improve phase modulation accuracy.

There are several possible routes for implementing the proposed tunable metasurfaces. By exploiting the thermo-optic effect, we could heat the nanoposts electrically or optically, and for the required change in refractive index, a temperature change of ~ 317 K would be necessary (see Supplementary Equation 20). This is slightly higher than the temperature generally used in silicon photonics, but could be achieved using silicon microheaters[59]. Alternatively, neglecting thermo-optic effects, we could also achieve tuning by injecting free carriers through photogeneration or forward biasing if we fabricate our nanoposts as p-i-n junctions (see Supplementary Equations 21-24 and Fig. S7 online). To achieve the necessary index modulation for the posts by free carrier refraction alone, we use a Drude model and calculate a required incident laser intensity of $1.26 \ MW/cm^2$ at a pump wavelength of 500 nm. If structured as a p-i-n junction, with posts as scatterers, individually addressing each element would be challenging as routing electrical traces to each pillar would be difficult due to the subwavelength lattice constant; however, our asymmetric device structure can be generalized to other scattering element geometries for which electrical routing would be simpler, such as 1-D unit cells of rectangles (see Supplementary Fig. S6 online). Another implementation route is to extend the design to other material platforms, such as phase-change materials, which can achieve unity-order changes in the refractive index via electrical[60] or optical[48] heating. Unlike previous implementations, one can pattern phase-change materials to create a tunable metasurface and also ensure electrical isolation between different scatterers. With phase-change material platforms, loss can be substantial depending on the operating wavelength, which requires careful design to ensure good performance (see Supplementary Equations 1-19 for the asymmetric Fabry-Perot equations incorporating loss).

For tuning methods based on optical excitation of the scatterers, we would need a spatially variant intensity function which could appropriately modulate refractive index as a function of position. For our Drude model calculated carrier density change at 500 nm excitation, a conventional liquid crystal (LC) spatial light modulator (SLM) could be used to produce a structured wavefront that could impinge on the metasurface, inducing refractive index changes related to the local intensity. This approach however would be speed-limited by the refresh rate of the LC SLM and would require an optical setup with macroscopic refractive optics, which would counter the benefit of compactness provided by the metasurface. As such, solutions based on electrically exciting the scatterers are more promising in terms of delivering benefits in size, weight, power, and speed. Fabrication of such a device would be extensive, requiring cointegration of electronics and photonics, with a high density of electrical traces required for individually addressing scatterers to achieve arbitrary pixel-by-pixel phase control. For an $M \times$

$N$ pixel array, the complexity of the required control circuit would be $O(M \times N)$ as each scatterer would need a separate control line. If instead a memory element were incorporated with each pixel and the control lines were assembled as a conventional crossbar architecture[61], then we could reduce the control complexity substantially to $O(M + N)$ as we could address pixels by the intersection of their row and column traces. With this approach, pixel columns would be updated in a time-sequential fashion, limiting the speed of the device relative to simultaneously addressing all pixels in parallel with separate control signals, although the speed could still greatly exceed that of an approach based on tuning with another SLM. While the CMOS compatibility of the silicon-based scatterers would facilitate cointegration of the photonics with control circuitry and conventional electronic memory cells, scatterers with our device structure based on phase-change memory media could also deliver their own unique benefits, with the possibility of achieving both the desired optical properties and memory storage capabilities simultaneously. With recent work[48] demonstrating grayscale changes in the optical properties of GeSbTe, phase-change materials could deliver analog refractive index control for inducing nonlinear phase shifts without having to constantly apply an external perturbation to maintain a scatterer's optical properties as the material would exist in a stable amorphous, crystalline, or intermediate state.

## Conclusion:

We reported an asymmetric Fabry-Perot-inspired tunable metasurface consisting of a high reflectivity bottom mirror with scattering nanoposts on top. While several implementations of such phase shifters exist, we report preserved cavity functionality even when our pixels consist of individual nanoposts, due to limited coupling between the elements. This enables subwavelength spatial resolution, and on-axis and in-plane focal scanning are possible even with phase curvatures high enough to achieve nearly diffraction-limited focusing at 100 µm; however, the element coupling is not weak enough to accurately realize higher gradient phase profiles including those of holograms or high order polynomial surfaces using the tunable metasurface. With the small index modulation range required, experimental implementation of tunable asymmetric elements is possible via either electrical biasing or optical excitation, the selection of which may depend on the designer's choice of scattering element geometry.

**Acknowledgements:**

This work was facilitated though the use of advanced computational, storage, and networking infrastructure provided by the Hyak supercomputer system at the University of Washington (UW). The research work is supported by the startup fund provided by the UW, Seattle, and the Intel Early Career Faculty Award. S.C. acknowledges support from the Paul C. Leach Fellowship from the UW Electrical Engineering department.

**Author Contributions:**

S.C and A.M conceived the design concept. S.C and A.Z performed the simulations. S.C wrote the manuscript. All authors discussed the results and commented on the manuscript.

**Competing Interests:**

The authors declare no competing financial interests.

**Figures:**

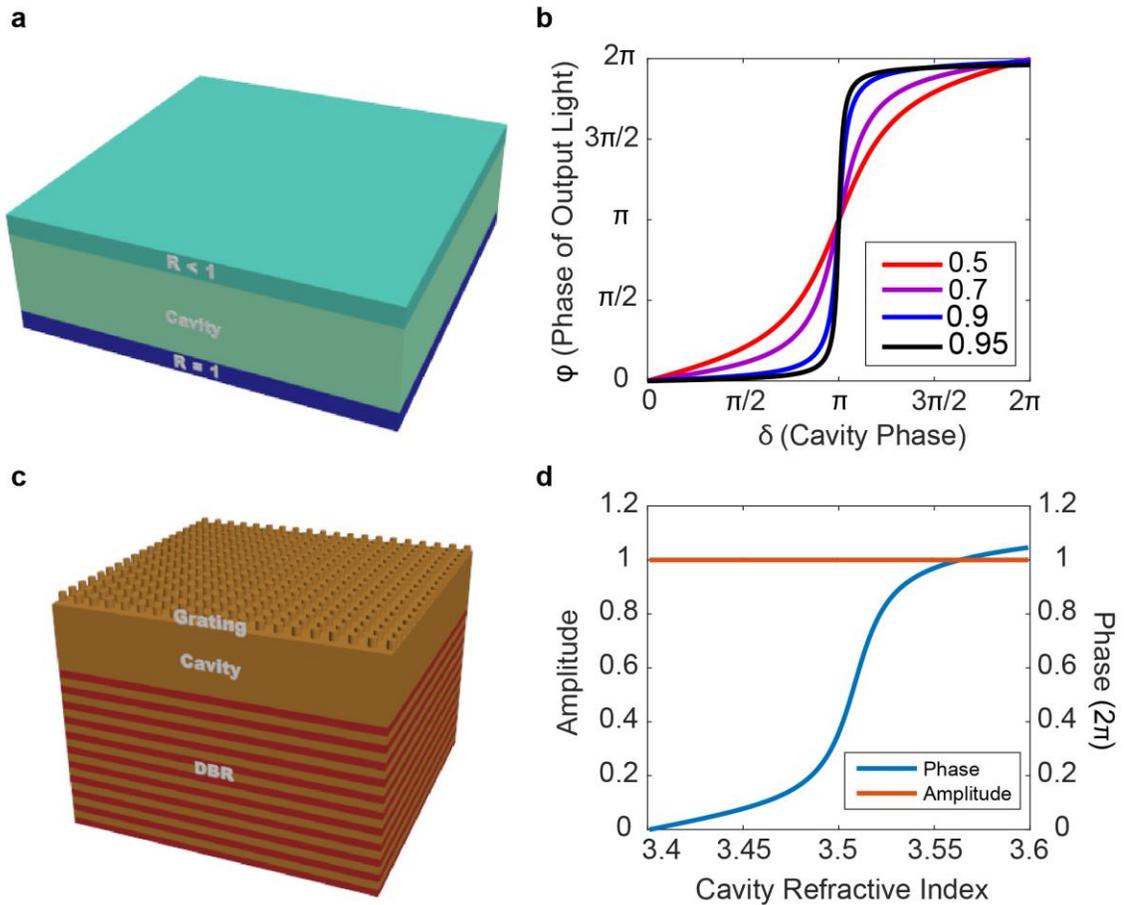

*Figure 1: Ideal asymmetric Fabry-Perot cavity-based phase shifter: (a) Schematic of the device with corresponding phase characteristics in (b) for different values of the top mirror reflectivity r. (c) Schematic of a realizable device with a DBR-based bottom mirror and high contrast grating top reflector. (d) RCWA-simulated phase characteristics for an example structure like that in (c) with parameters found in the text.*

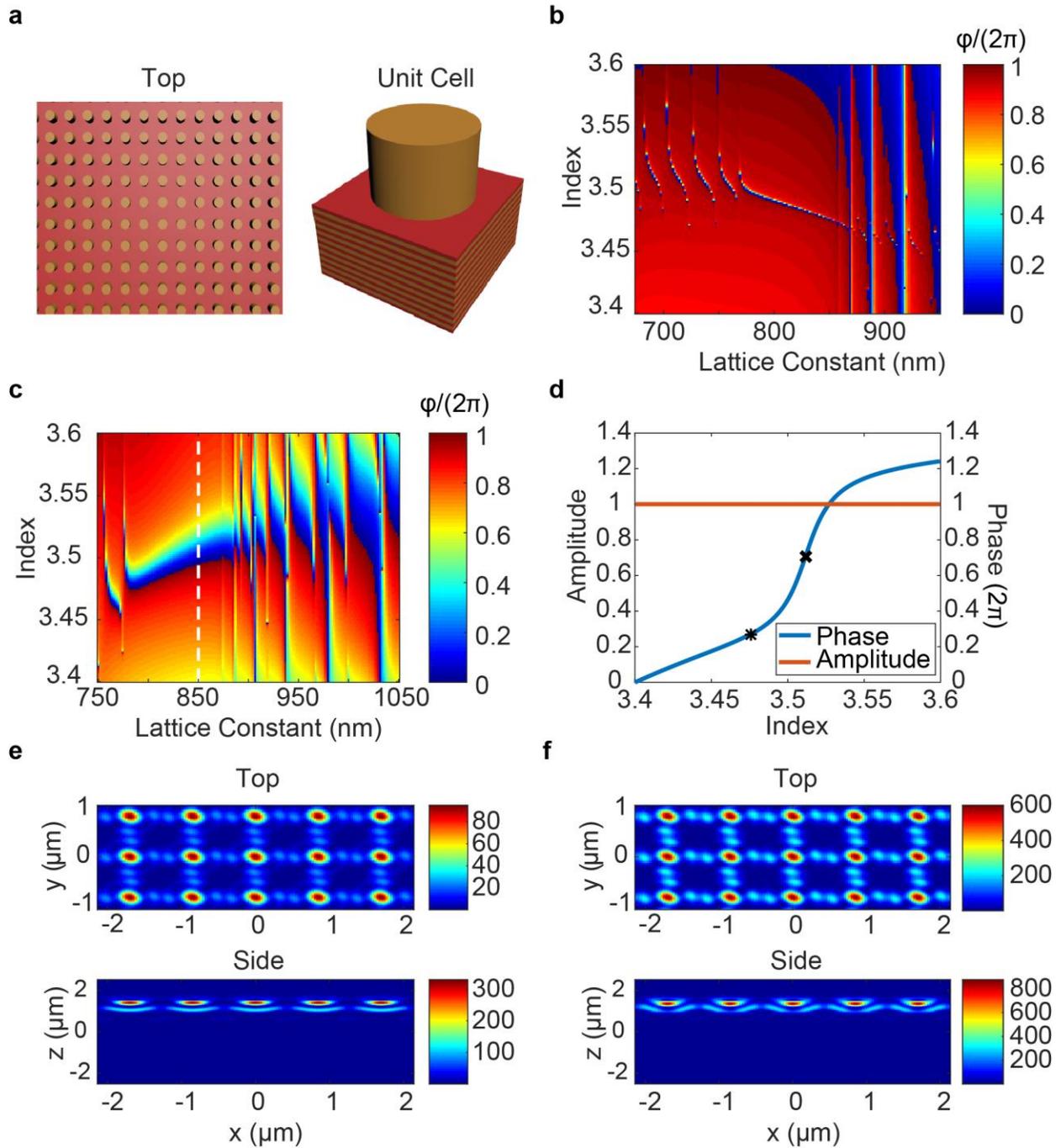

*Figure 2: Design of scattering nanoposts: (a) Top view schematic of a metasurface composed of nanoposts atop a DBR with the unit cell shown, RCWA-calculated reflection coefficients as a function of index and lattice constant for rapid (b) and moderate (c) phase change regimes, (d) reflection coefficient for a fixed period corresponding to the white dashed line in (c) with phase*

*adjusted to not wrap modulo 2π, and magnetic energy density profiles for when the incident wave has a density of unity for off (e) and on (f) resonance cases, corresponding to refractive indices indicated by the * and X in (d) respectively.*

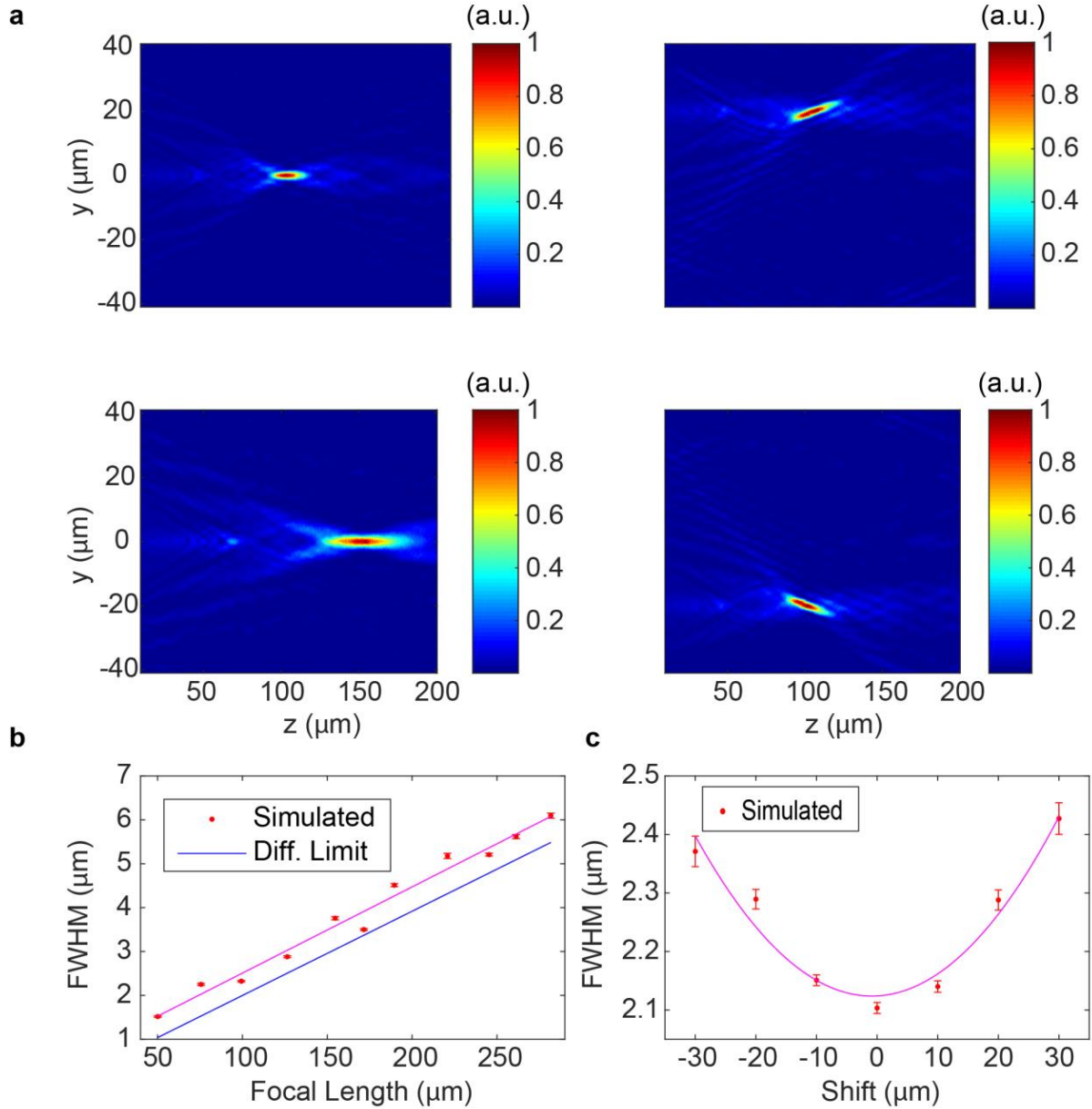

*Figure 3: Tunable aspherical lenses: (a) FDTD-simulated intensity profiles for focal scanning on-axis (left) and in-plane (right) (b) Spot size as a function of focal length (c) Spot size as a*

*function of in-plane shift. The magenta lines are eye guides and the error bars give the 95% confidence interval derived from fitting error.*

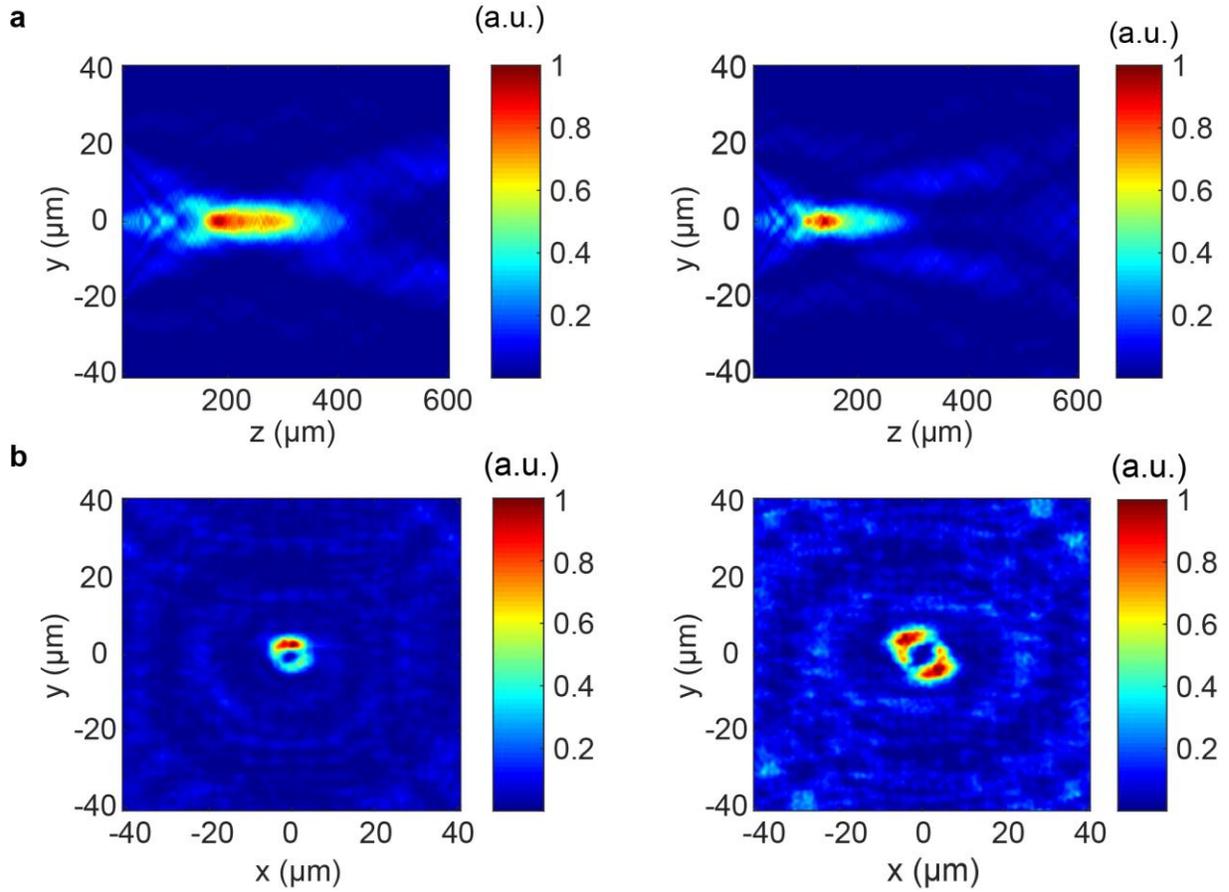

*Figure 4: Tunable metasurface axicons and vortex beam generators: (a) Intensity profiles for axicons with $\beta = 4°$ (left) and $\beta = 5°$ (right) (b) Intensity profiles for a vortex beam generator with $l = 1$ (left) and $l = 2$ (right)*